\documentclass[10pt,conference]{IEEEtran}
\IEEEoverridecommandlockouts
\usepackage{cite}
\usepackage{amsmath,amssymb,amsfonts}
\usepackage{algorithmic}
\usepackage{graphicx}
\usepackage{textcomp}
\usepackage{xcolor}
\usepackage{booktabs}
\def\BibTeX{{\rm B\kern-.05em{\sc i\kern-.025em b}\kern-.08em
    T\kern-.1667em\lower.7ex\hbox{E}\kern-.125emX}}
\begin{document}

\title{Are We Aligned? A Preliminary Investigation of the Alignment of Responsible AI Values between LLMs and Human Judgment}

\author{
Asma Yamani\textsuperscript{1},
Malak Baslyman\textsuperscript{1,2},
Moataz Ahmed\textsuperscript{1,3} \\
\textsuperscript{1}Information and Computer Science Department, KFUPM, Dhahran, Saudi Arabia \\
\textsuperscript{2}IRC for finance and digital economy, KFUPM, Dhahran, Saudi Arabia \\
\textsuperscript{3}SDAIA-KFUPM Joint Research Center for Artificial Intelligence, KFUPM, Dhahran, Saudi Arabia \\
\{g201906630, malak.baslyman, moataz\}@kfupm.edu.sa
}

\maketitle

\begin{abstract}
Large Language Models (LLMs) are increasingly employed in software engineering tasks such as requirements elicitation, design, and evaluation, raising critical questions regarding their alignment with human judgments on responsible AI values. This study investigates how closely LLMs’ value preferences align with those of two human groups—a US-representative sample and AI practitioners. We evaluate 23 LLMs across four tasks: (T1) selecting key responsible AI values, (T2) rating their importance in specific contexts, (T3) resolving trade-offs between competing values, and (T4) prioritizing software requirements that embody those values. The results show that LLMs generally align more closely with AI practitioners than with the US-representative sample, emphasizing fairness, privacy, transparency, safety, and accountability. However, inconsistencies appear between the values that LLMs claim to uphold (Tasks 1–3) and the way they prioritize requirements (Task 4), revealing gaps in faithfulness between stated and applied behavior. These findings highlight the practical risk of relying on LLMs in requirements engineering without human oversight and motivate the need for systematic approaches to benchmark, interpret, and monitor value alignment in AI-assisted software development.
\end{abstract}


\section{Introduction}
The rise of AI-powered software creation is transforming how requirements are gathered, values are embedded, and systems are governed. As Large Language Models (LLMs) and multi-agent AI systems increasingly participate in co-developing software systems~\cite{jin2024llmsllmbasedagentssoftware,10628428,rasnayaka2024empiricalstudyusageperceptions,10.1007/978-3-031-62110-9_1}, as well as automated machine learning (AutoML) systems~\cite{trirat2024automlagentmultiagentllmframework,xu2024largelanguagemodelssynergize,tornede2024automlagelargelanguage}, a critical question arises: \emph{Which responsible AI values do these models encode?}

Developing responsible AI-based systems requires adopting and embedding a clear set of values throughout the entire software lifecycle, with fairness, explainability, and accountability at its core~\cite{BARREDOARRIETA202082}. While responsible AI values may vary by organization, the Berkman Klein Center identified eight primary themes under which these values are categorized: privacy, accountability, safety, transparency, fairness, human control, professional responsibility, and the promotion of human values~\cite{Fjeld2020}. Recent efforts, such as the MALEA framework for ethics-aware requirements elicitation, illustrate how multi-agent LLM systems can generate or critique ethical requirements early in the software lifecycle, facilitating the embedding of responsible AI values into AI systems~\cite{yamani2025multiagentllmsethicsadvocates}. Also, the work in~\cite{decerqueira2025trustaiagentscase} aims to evaluate code compliance with ethical standards, providing recommendations to
embed trustworthiness from early stages of development. However, a study by Jakesch et al.~\cite{Jakesch2022} revealed that AI practitioners and end-users (represented by a US census representative sample) perceive the importance of responsible AI values differently. This divergence raises critical concerns about \emph{how differing responsible AI values transfer into LLMs, potentially resulting in value misalignment within AI systems co-developed by LLMs}. 

In this study, we empirically investigate the alignment between LLMs and human judgments regarding responsible AI values. We use the term alignment to refer specifically to the statistical correlation between the rankings or selections of Responsible AI values produced by LLMs and those reported by human groups—namely, AI practitioners and a US-census representative sample. Moreover, we explore whether alignment (or lack thereof) observed in stated value preferences translates into practical software engineering decisions. Specifically, we aim to:
\begin{itemize}
    \item Evaluate the alignment between LLMs and both the US-representative sample and AI practitioners in selecting important responsible AI values.
    \item Examine alignment between LLMs and both human groups when rating the importance of responsible AI values across four different application contexts.
    \item Investigate alignment in resolving trade-offs among responsible AI values within these contexts.
    \item Assess the consistency between the value rankings in trade-off resolutions and outcomes in practical software requirements prioritization tasks.
\end{itemize}

We achieve these objectives by empirically evaluating 15 LLMs of varying sizes and architectures, applying the study design proposed initially for human subjects by Jakesch et al.~\cite{Jakesch2022}. Our results indicate that LLMs generally align more closely with AI practitioners than the US-representative sample across abstract value assessments, suggesting practitioner influence in model engineering processes. Furthermore, we introduce a novel fourth task, requirements prioritization, simulating realistic decision-making scenarios in software engineering. By comparing the prioritization outcomes of value-related user stories to the stated value preferences in previous tasks, we observe significant inconsistencies. While alignment with AI practitioners exists for some values, it does not consistently translate into actual prioritization behaviors. This highlights the need for deeper alignment processes to ensure consistency and reliability in AI-driven software engineering tasks.

\section{Related Work}
Measuring the moral compass of LLMs has recently become a focus of multiple studies, with their widespread use in various applications~\cite {ji2024moralbenchmoralevaluationllms, leaderboard, Norhashim_Hahn_2024}.
Ji et al.~\cite{ji2024moralbenchmoralevaluationllms} introduced a new benchmark specifically designed to evaluate the moral identity of LLMs. The benchmark questions are based on the moral foundation theory, which covers six aspects: Care, Fairness, Loyalty, Authority, Sanctity, and Liberty. The benchmark included 132 scenarios, and the analysis provided insight into the underlying mechanisms influencing the LLMs' responses. Norhashim and Hahn~\cite{Norhashim_Hahn_2024} prompted GPT-3.5 repeatedly with scenarios from the Moral Stories dataset~\cite{emelin2020moralstoriessituatedreasoning} to investigate the consistency of values and alignment with responsible AI values. Moreover, it classified norms from the Moral Stories dataset into broader abstract values to aid developers in identifying and prioritizing areas for enhancement. Ye et al.~\cite{ye2025measuringhumanaivalues} proposed the Generative Psychometrics for Values metric, a novel LLM-based value measurement paradigm grounded in text-revealed selective perceptions. They fine-tuned the Llama3-8B model to detect such values, introducing ValueLama and using it for extensive evaluations across 17 LLMs and evaluating human-written blogs. 

Developing a benchmark that can demonstrate that an LLM is fully ethical is impossible, as there is no ground truth to moral dilemmas. Additionally, the ethics of AI is a long-tail problem, and it is impossible to represent every contingency in the training data and test every outcome produced by the models. An approach to mitigate this challenge is to measure the alignment of values based on specific definitions in a certain context~\cite{lacroix2022metaethicalperspectivesbenchmarkingai}.
Gomez-Vazquez et al.~\cite{leaderboard} released a public leaderboard that evaluates $16$ LLMs on over $300$ hundred input tests, spanning seven different social biases. In the context of software engineering, Morales et al.~\cite{10.1145/3640310.3674093} proposed LangBiTe, a model-driven solution to specify ethical requirements and customize and automate testing ethical biases in LLM. In their framework, requirements engineers select which ethical concerns to test if the LLM is biased against and the sensitive communities targeted for each concern. Then, test cases are generated, and the evaluation is executed to identify potential biases. \par

As LLMs continue to emerge as tools to assist in software development and as components in AutoML systems, LLMs' views on responsible AI values will begin to have moral weight and implications for the software they helped design. Therefore, it is essential to investigate LLMs' judgment regarding responsible AI values. Hence, in this work, we examine whether LLMs are aligned with human judgment regarding ethical values for responsible AI and explore how the reported responsible AI profile translates into the practical application of software requirements prioritization.

\section{Study Design}

\textbf{Procedure} To ensure that the LLMs have some background on the relevant concepts, we first asked each model to define each responsible AI value. We planned to exclude any LLM lacking knowledge of these concepts. However, all LLMs provided definitions aligned with the concepts described by Jobin et al.~\cite{Jobin2019}.  Next, we conducted a preliminary investigation with 23 LLMs to assess their robustness when selecting AI values. Details of the robustness evaluation procedure—including standard deviation and Jaccard-similarity computations across 150 runs—are provided in the supplementary material.~\footnote[1]{https://github.com/asmayamani/ResponsibleAI\_LLM\_Alignment}.
Following this preliminary assessment of 23 LLMs, we selected 15 models that exhibited above-average robustness for the first three main tasks . From these, a subset of eight LLMs that achieved perfect Jaccard similarity and above-average robustness were included in the in-depth analysis and the novel fourth task. Results for the remaining seven LLMs, which showed moderate robustness, are provided in the supplementary material for benchmarking~\footnotemark[1]. This hierarchical selection ensured that later analyses focused on the most consistent and reproducible LLMs.
To investigate how well the LLMs' values align with those of the US population and AI practitioners, we followed the study design from~\cite{Jakesch2022}, which involved three main tasks: \textbf{Task 1 (T1):} Identifying the most important values in AI ethics by selecting the top 5 values out of 12 values. \textbf{Task 2 (T2):} Rating the importance of responsible AI values across four different contexts. \textbf{Task 3 (T3):} Prioritizing specific responsible AI values that might conflict during development. While the original study by Jakesch et al.~\cite{Jakesch2022} involved human participants, our study examines the most robust LLMs identified during our preliminary assessment. Alignment is operationalized through Spearman’s $\rho$ between model- and human-derived value rankings across these tasks.

We then introduced a novel fourth task to examine whether the stated preferences in Tasks 1-3 translate into practical software engineering decisions through a realistic scenario of requirements prioritization, mirroring real-world software engineering contexts. This is crucial given the recent concerns highlighted by Agarwal et al. \cite{agarwal2024faithfulnessvsplausibilityunreliability} and Turpin et al. \cite{turpin2023languagemodelsdontsay} regarding the potential misalignment between LLM-stated preferences and their actual behavior. 
\textbf{Task 4 (T4):} Prompting LLMs to prioritize high-level user story requirements. The comparative analysis between tasks T2, T3 and T4 sheds light on whether the stated importance of ethical principles influences actual behavior when applied to a practical example in software engineering, thus addressing the concerns around LLM faithfulness. This fourth task specifically challenges LLMs to practically integrate ethical values into realistic constraints and trade-offs typical in software development scenarios. 

\textbf{Values.} To perform a competitive analysis with the human reported value inclinations in~\cite{Jakesch2022} we evaluate the LLMs for the same values. For Task T1, the values included are the AI ethical principles listed in ~\cite{Jobin2019}, which include the following \texttt{responsible AI values~\cite{Jakesch2022}}: Performance, Privacy, Fairness, Human Autonomy, Accountability, Safety, and Transparency. It also included additional values such as Inclusiveness, Dignity, Social Good, Sustainability, and Solidarity. Task 2 included the \texttt{responsible AI values}. In Task T3, the value conflicts we used the same set examined by~\cite{Jakesch2022} drawn from the literature which include: Performance vs. Privacy~\cite{bagdasaryan2019differentialprivacydisparateimpact,Shokri2015}, Fairness vs. Privacy~\cite{bagdasaryan2019differentialprivacydisparateimpact, pmlr-v81-ekstrand18a}, Fairness vs. Performance~\cite{CorbettDavies2017, pleiss2017fairnesscalibration}, Autonomy vs. Safety~\cite{Livingstone2011RisksSafetyInternet}, Transparency vs. Safety~\cite{hua2021increasingadversarialuncertaintyscale,10.1145/1774088.1774151}.As for Task T4, the values considered were the same values investigated in T3: Performance, Privacy, Fairness, Autonomy, Safety, and Transparency. 

\textbf{Contexts and requirements.} The four contexts used in this study include: Medical, Banking, Marketing, and Streaming contexts. 
As there are no established software engineering examples for prioritizing responsible AI values related requirements, we created a synthetic set for Task 4 using o1-mini to generate requirements from four abstracts~\cite{10.1145/3727582.3728689} of existing real-world machine learning applications in the same four contexts: (1) A glucose level monitoring system for diabetes management~\cite{plis2014machine} (Medical context), (2) A credit card default prediction system~\cite{sayjadah2018credit} (Banking context), (3) A music playlist success prediction system~\cite{10454829} (Streaming context), (4) An advertisement placement prediction system for internet pages~\cite{10602683} (Marketing context). After generation, the requirements were edited by the first author and reviewed by the second and third authors, who are academics with industry experience. The requirements are available in our online repository~\footnotemark[1].

\textbf{LLMs.} The preliminary investigation, detailed in our online repository\footnotemark[1], categorized the 23 LLMs into three groups based on their robustness in selecting AI values. 

\textbf{Group 1} (8 LLMs) comprised the most robust models that achieved perfect Jaccard similarity and above-average robustness; these were analyzed in all tasks (T1–T4). This group includes \{Gemini-1.5-flash~\cite{geminiteam2024geminifamilyhighlycapable}, Gemini-2.0-flash-exp, O1-mini, O1-preview, Llama 3.1 405b~\cite{touvron2023llama}, WizardLM-2-8x22B~\cite{xu2024wizardlm2}, DeepSeek-v3~\cite{deepseekai2025deepseekv3technicalreport}, and Phi-4~\cite{abdin2024phi4technicalreport}\}.

\textbf{Group 2} (7 LLMs) demonstrated moderate robustness and were evaluated in T1–T3; their detailed results appear in the supplementary material for completeness.\footnotemark[1] This group includes \{GPT-4o~\cite{openai2024gpt4technicalreport}, GPT-4o-mini, Gemini-1.5-pro, Gemma 27B~\cite{gemmateam2024gemma2improvingopen}, Claude-3.5-Sonnet~\cite{anthropic2024claude3}, Llama 3.3 70B, and Llama 3.1 70B\}.

\textbf{Group 3} (8 LLMs) showed low robustness and were included only in the value-selection task (T1) for comparison.\footnotemark[1] This group includes \{QwQ-32B-Preview, Qwen2.5-72B-Instruct~\cite{bai2024qwen2}, Llama 3.1 8B, Mixtral-8x7B-Instruct-v0.1~\cite{jiang2024mixtral}, Llama 3.2 3B, Claude-3.5-Haiku, Gemma 9B, and Gemma 2B\}. It is worth noting that no LLM was excluded for lack of knowledge of the values; subsequent analyses concentrated on the most stable and reproducible models.

\textbf{Prompts and Environment} The prompts used in the study are provided in our online repository~\footnotemark[1]. Each prompt was executed 50 times for each LLM, with the cache cleared before each run to account for any randomness in the outputs. We used the Deepinfra API for open-source models and the respective APIs for proprietary models. All hyperparameters were kept at their default settings, except for the temperature, which was set to $1.0$. We set this temperature value to capture a wide range of possible responses from each LLM and account for the variability in the models' outputs, which is essential for accurately assessing the LLM alignment across multiple runs. 

To ensure the robustness of our results, we randomized the choices of values, rankings, and importance levels in each run. Additionally, we conducted a synonym robustness test for T1 by substituting original principle names with synonymous or closely related terms (e.g., alternating "Fairness," "Justice," "Non-discrimination"; "Performance," "Accuracy"; "Privacy," "Data Protection"). Full list of synonyms is in our online repository~\footnotemark[1]. Moreover, to mitigate any ordering effects in the prompts for Task T3, we presented the value conflicts in both directions. For example, we asked the LLMs to consider "ensuring privacy while reducing fairness" as well as "ensuring fairness while reducing privacy." 

\textbf{Aggregation of LLM Outputs}
To systematically analyze the outputs generated by the LLMs, we employed Python's difflib library\footnote{https://docs.python.org/3/library/difflib.html} to robustly map each model-generated response to its corresponding choice, handling variations such as punctuation, additional whitespace, capitalization discrepancies, and response repetitions. Because LLMs selected from fixed choices, text differences were syntactic; semantic similarity checks were unnecessary. The mapped responses were then aggregated by counting to compute the percentage distribution for each choice across all runs. To enhance transparency and reproducibility, Python code is publicly available in our online repository~\footnotemark[1].

\textbf{Human Judgment Collection}
Human judgment data used for comparison with LLM outputs was directly obtained from the published figures in \cite{Jakesch2022}. Their research employed stratified random sampling, to match the distribution closely to U.S. census demographics, ensuring representativeness. They also performed attention checks during the survey to ensure the validity of the responses. The total number of valid responses for the US census representative sample was 607, while for AI practitioners it was 140.

\section{Results}
In this section, we report the detailed results for the 15 LLMs included in the main alignment tasks (T1–T3), highlighting the eight most robust models used in T4; results for the remaining seven LLMs with moderate robustness are provided in the supplementary material. Also numerical results and standard deviation for benchmarking~\footnotemark[1].\par
\subsection{What values are reported as most important by LLMs?}
\begin{figure}
    \centering
    \includegraphics[width=\linewidth]{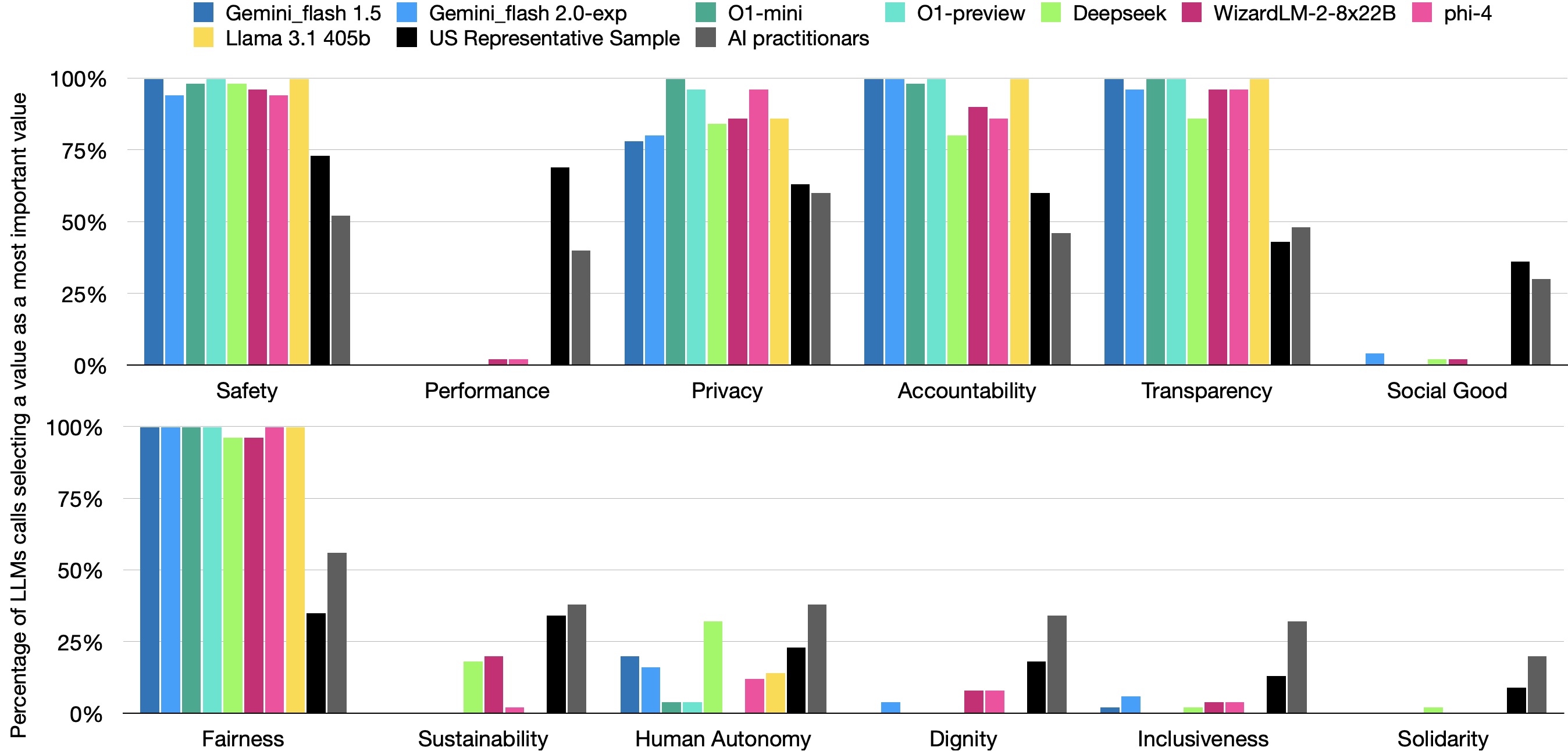}
    \caption{Results of selecting the most important five values from 12 values. LLMs align with AI practitioners more than the US-representative sample selection. $N_{per LLM} = 50$, $N_{US-representative sample} = 516$, $N_{AI practitioners} = 140$.}
    \label{fig:Q1}
\end{figure}

We probed LLMs to select the top five ethical principles without considering any specific context. The findings indicate that the LLMs in Group 1 are fully aligned with AI practitioners regarding the five most important values: fairness, privacy, transparency, safety, and accountability, achieving a Jaccard Similarity of 1.0. Specifically, fairness was identified as one of the most important values in 99\% of LLM runs. Safety, transparency and accountability were selected in 98\%, 97\%, and 94\% of the runs, respectively, while privacy was included in 88\% of runs. Additionally, human autonomy emerged as the sixth most important value, selected among the top five in 12\% of LLM runs, aligning with AI practitioners.

When compared to a US-representative sample, the LLMs achieved a Jaccard Similarity of 0.67, since the models prioritize fairness over performance in this selection task. Furthermore, as shown in Table \ref{T1}, when treating the selection as a ranking problem based on the percentage of selections, the analysis reveals that all LLMs have a moderately positive correlation with the US-representative sample. However, these correlations are not statistically significant. In contrast, the LLMs demonstrate a statistically significant strong positive correlation with the responses from AI practitioners, as detailed in Table \ref{T1}.

Regarding sensitivity to the prompt, more specifically, to the principles' terms used in the prompt, the synonyms robustness experiment revealed only minor frequency shifts in principle selection; notably, using the term "accuracy" slightly elevated the importance of "performance." However, the top 5 values remained identical for 7 of 8 LLMs (Jaccard similarity=1.0) based on aggregated values of 50 runs. The 8th LLM, Gemini-1.5-flash, alternated between Human Autonomy (originally 6th) and Privacy (originally 5th).

\begin{table}[htbp]
\centering
\caption{Spearman’s correlation values comparing the selection of top ethical principles by LLMs with those prioritized by a US-representative sample and AI practitioners. \small{(Note: *p < 0.05, **p < 0.01, ***p < 0.001) \label{T1}}}
\begin{tabular}{p{0.3\columnwidth}p{0.31\columnwidth}p{0.25\columnwidth}}
\toprule
LLM             & US sample & AI Practitioners \\ \midrule
Gemini\_flash 1.5 & 0.455                                     & 0.763**                           \\
Gemini\_flash 2.0 & 0.371                                     & 0.694*                            \\
O1-mini           & 0.49                                      & 0.871***                          \\
O1-preview        & 0.533                                     & 0.803**                           \\
Deepseek          & 0.452                                     & 0.763**                           \\
WizardLM-2-8x22B  & 0.519                                     & 0.777**                           \\
phi-4             & 0.408                                     & 0.882***                          \\
Llama 3.1 405b    & 0.533                                     & 0.803**                           \\ \bottomrule
\end{tabular}

\end{table}
\subsection{How important are values in specific context?}
\begin{figure*}
    \centering
    \includegraphics[width=0.65\linewidth]{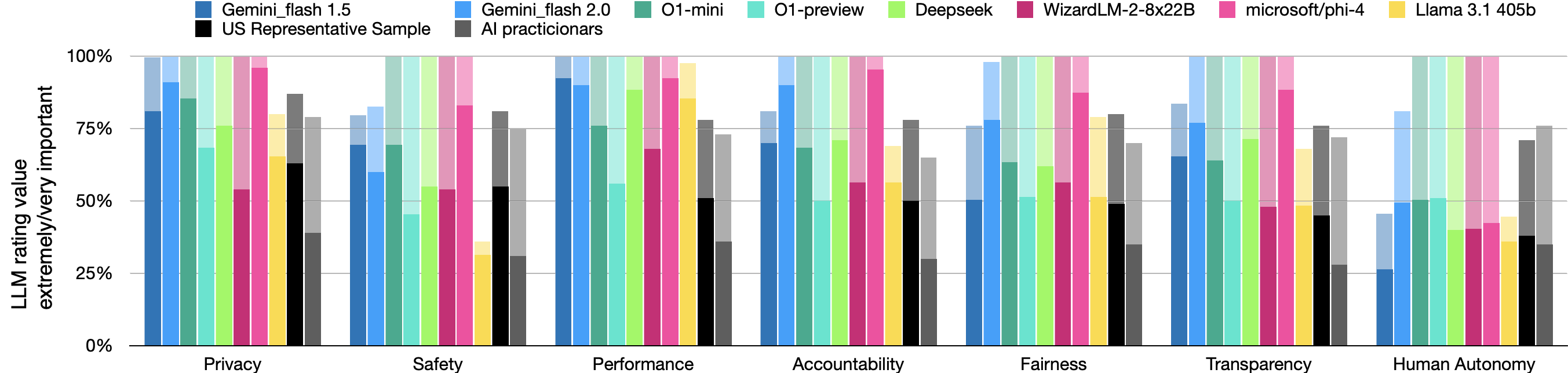}
    \caption{Percentage of LLM responses selecting a responsible AI value as Extremely important (dark color) or very important (light color) across the four investigated contexts. \small{$N_{per value,LLM} = 200$, $N_{US-representative sample and AI practitioners} = 140 - 607$.}}
    \label{fig:Q2byLLM}
\end{figure*}

We prompted LLMs to rate the importance of responsible AI principles across various contexts. As illustrated in Figure~\ref{fig:Q2byLLM}, LLMs generally assigned a higher importance to responsible AI values when compared to both AI practitioners and the US-representative sample. Among the AI-related values, performance was considered the most important by LLMs, followed by privacy and accountability, with human autonomy and safety ranking lowest and second lowest, respectively.

This prioritization differs from the responses of AI practitioners and the US-representative sample as for both groups, privacy emerged as the most valued principle. In terms of extremely important ratings, after privacy, the US-representative sample prioritized safety then performance, whereas AI practitioners prioritized fairness and human autonomy then performance.
Among the LLMs, Llama 3.1 405b had the lowest average proportion of responses rating values as very or extremely important. In contrast, models of O1-mini, O1-preview, Deepseek, WizardLM-2-8x22B, and microsoft/phi-4 consistently rated all values as very or extremely important.

\begin{figure*}
    \centering
    \includegraphics[width=0.65\linewidth]{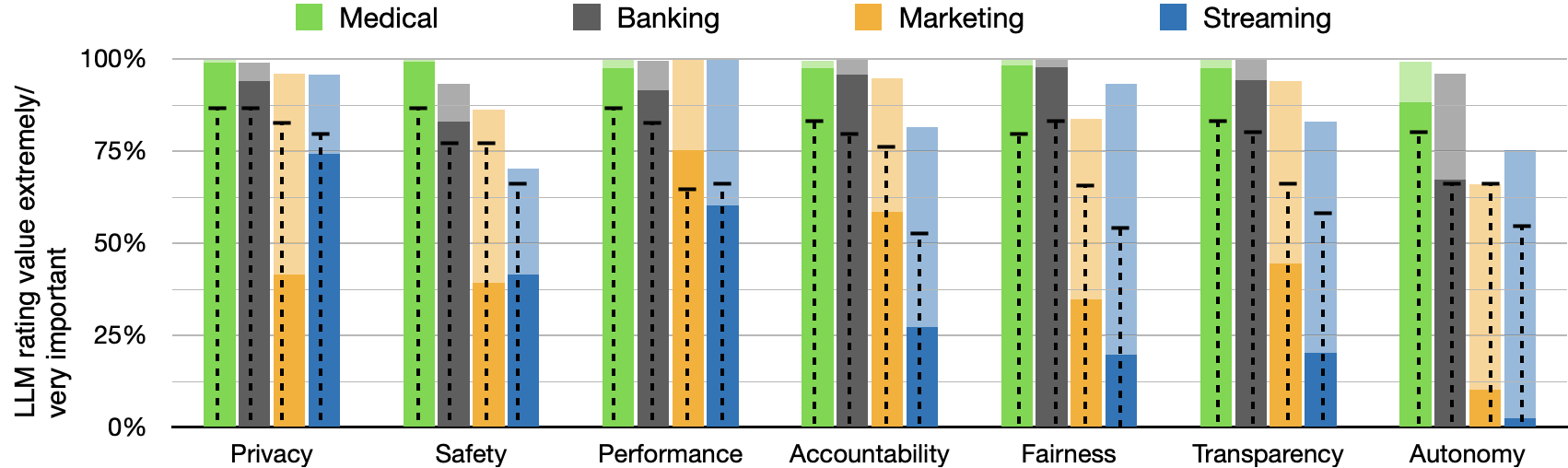}
    \caption{Percentage of LLM responses selecting a responsible AI value as Extremely important (dark color) or very important (light color) across the four investigated contexts. \small{$N_{per value,context} = 400$. The dotted bars illustrates the US-representative sample responding by extremely/very important to a certain value.}}
    \label{fig:Q2byContext}
\end{figure*}

Figure~\ref{fig:Q2byContext} presents the importance of values across different contexts. In the medical context, responsible AI values were mostly rated as  'extremely important' by LLMs, followed by the banking context. However, significant variability was observed among LLMs in the marketing and streaming contexts. Specifically, LLMs did not consider human autonomy as 'extremely important' in either the marketing or streaming contexts. In marketing, safety was rated second to last, followed by fairness, whereas in streaming, fairness was second to last, followed by safety. For performance, privacy, and accountability, over 50\% of responses were rated as extremely important on average across contexts, while in the streaming context, only performance surpassed this threshold.

The Spearman correlation coefficients between LLMs in Group 1 and the percentage of responses per (context, value) pair are presented in Table~\ref{T2}. All LLMs (except O1-preview, Llama 3.1 405b, and Gemini flash 1.5) exhibited a statistically significant strong monotonic positive correlation with the US-representative sample. Moreover, O1-mini, O1-preview, and WizardLM-2-8x22B, and Llama 3.1 405b demonstrated a statistically significant strong monotonic positive correlation with AI practitioners. Overall, LLMs aligned more closely with AI practitioners than with the US-representative sample. Additionally, LLMs showed a moderate negative Spearman correlation of -0.536 (p-value = 0.04) between their correlations with AI practitioners and the US-representative sample when considering both Group 1 and Group 2\footnote{This relationship also holds for Group 1 alone, although it is not statistically significant.}. This indicates that LLMs that are more aligned with AI practitioners are less aligned with the US-representative sample and vice versa.

\begin{table}[htbp]
\centering
\caption{Spearman’s correlation values for LLMs ranking the importance of a value per context with respect to the US-representative sample and AI Practitioners.\small{ (Note: *p < 0.05, **p < 0.01, ***p < 0.001) \label{T2}}}

\begin{tabular}{p{0.3\columnwidth}p{0.31\columnwidth}p{0.25\columnwidth}}
\toprule
LLM             & US sample & AI Practitioners \\ \midrule
Gemini\_flash 1.5 & 0.661***                                                      & 0.696***                                              \\
Gemini\_flash 2.0 & {0.763***}                                       & 0.630***                                              \\
O1-mini           & { 0.747***}                                                & {0.768***}                           \\
O1-preview        & 0.588***                                                      & {0.785***}                             \\
Deepseek          & { 0.719***}                                                & 0.654***                                              \\
WizardLM-2-8x22B  & { 0.705***}                                                & { 0.714***}                                        \\
microsoft/phi-4   & { 0.739***}                                                & 0.479**                                               \\
Llama 3.1 405b    & 0.654***                                                      & { 0.711***}                                        \\ \bottomrule
\end{tabular}

\end{table} 
\subsection{How are value trade-offs resolved?} 
\begin{figure}
    \centering
    \includegraphics[width=\linewidth]{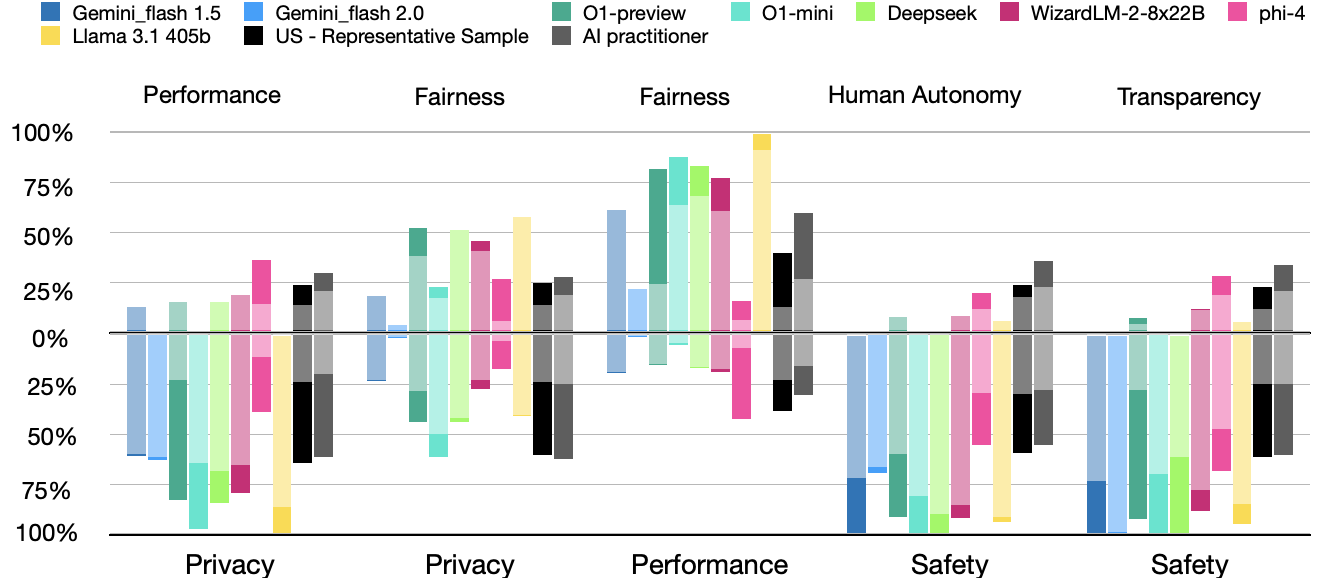}
    \caption{Percentage of LLM responses prioritizing a certain value over another across the four contexts. \small{Lightly shaded area is a lower degree of preference. Undecided values are omitted. $N_{per value,LLM} = 200$, $N_{US-representative sample and AI practitioners} = 140 - 607$}}
    \label{fig:Q3a}
\end{figure}

We prompted LLMs to prioritize one value over another when faced with conflicts within specific contexts. As shown in Figure~\ref{fig:Q3a}, LLMs consistently favored safety over both human autonomy and transparency. Additionally, LLMs generally prioritized privacy and fairness above performance. An exception to this pattern was observed with the WizardLM-2-8x22B model, which favored performance over fairness in multiple instances. However, in the case of fairness-privacy trade-off. The Gemini-flash models and phi-4 exhibited a high level of indecision, with more than 50\% of responses indicating no clear preference. Similarly, the Gemini-2.0-flash model showed a significant indecision in balancing fairness with performance. Detailed visualization of value trade-off resolution per LLM and context are in our online repository~\footnotemark[1].

\begin{figure}
    \centering
    \includegraphics[width=\linewidth]{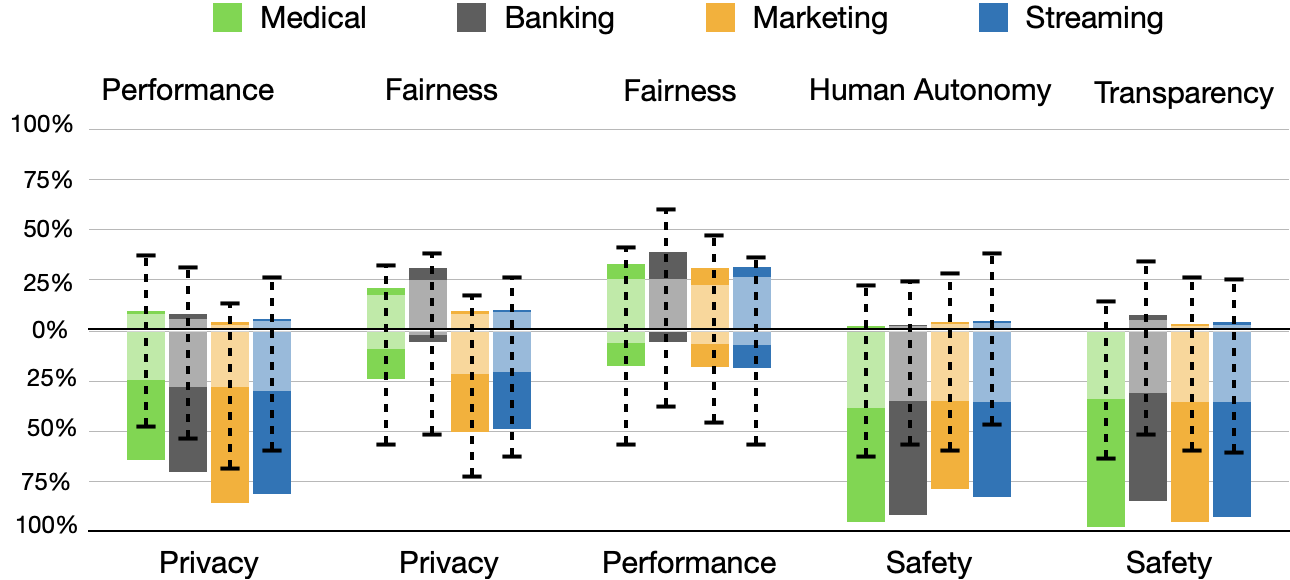}
    \caption{Percentage of LLM responses prioritizing a certain value over another by context. \small{Lightly shaded area is a lower degree of preference.  Undecided values are omitted. $N_{per value,context} = 400$. The dotted bars illustrates the US-representative sample responding by strongly/somewhat prioritizing a certain value.}}
    \label{fig:Q3b}
\end{figure}

On average, LLMs demonstrated a slightly higher alignment with AI practitioners compared to the US-representative sample in terms of Spearman correlation. Nevertheless, given that AI practitioners and the US-representative sample were themselves highly aligned in this task, thus LLMs exhibited a very strong Spearman correlation with both groups, all with high statistical significance. Detailed Spearman correlation values and p-values are presented in Table~\ref{T3}. Figure~\ref{fig:Q3b} illustrates how values were prioritized across different contexts. In the banking context, LLMs were slightly less inclined to prioritize safety over transparency compared to other contexts. Similarly, in the marketing and streaming contexts, LLMs were marginally less decisive in favoring safety over human autonomy. Furthermore, fairness was slightly more favored over performance in the banking context than in others. Additionally, fairness was somewhat more prioritized over privacy in both the banking and medical contexts compared to other settings.

\begin{table}[htbp]
\caption{Spearman’s correlation values for LLMs prioritizing a certain value over another by context with respect to the US-representative sample and AI Practitioners. \small{(Note: *p < 0.05, **p < 0.01, ***p < 0.001)\label{T3}}}
\centering
\begin{tabular}{p{0.3\columnwidth}p{0.31\columnwidth}p{0.25\columnwidth}}
\toprule
LLM             & US sample & AI Practitioners \\ \midrule
Gemini\_flash 1.5 & 0.850*                   & 0.859*                         \\
Gemini\_flash 2.0 & 0.852*                   & 0.877*                         \\
O1-preview        & 0.884*                   & 0.854*                         \\
O1-mini           & 0.823*                   & 0.869*                         \\
Deepseek          & 0.844*                   & 0.859*                         \\
WizardLM-2-8x22B  & 0.835*                   & 0.866*                         \\
Microsoft/phi-4   & 0.805*                  & 0.817*                         \\
Llama 3.1 405b    & 0.835*                   & 0.872*                         \\ \bottomrule
\end{tabular}
\end{table}

\subsection{How Do LLMs Prioritize High-Level Requirements?}
\begin{figure*} \centering \includegraphics[width=0.7\linewidth]{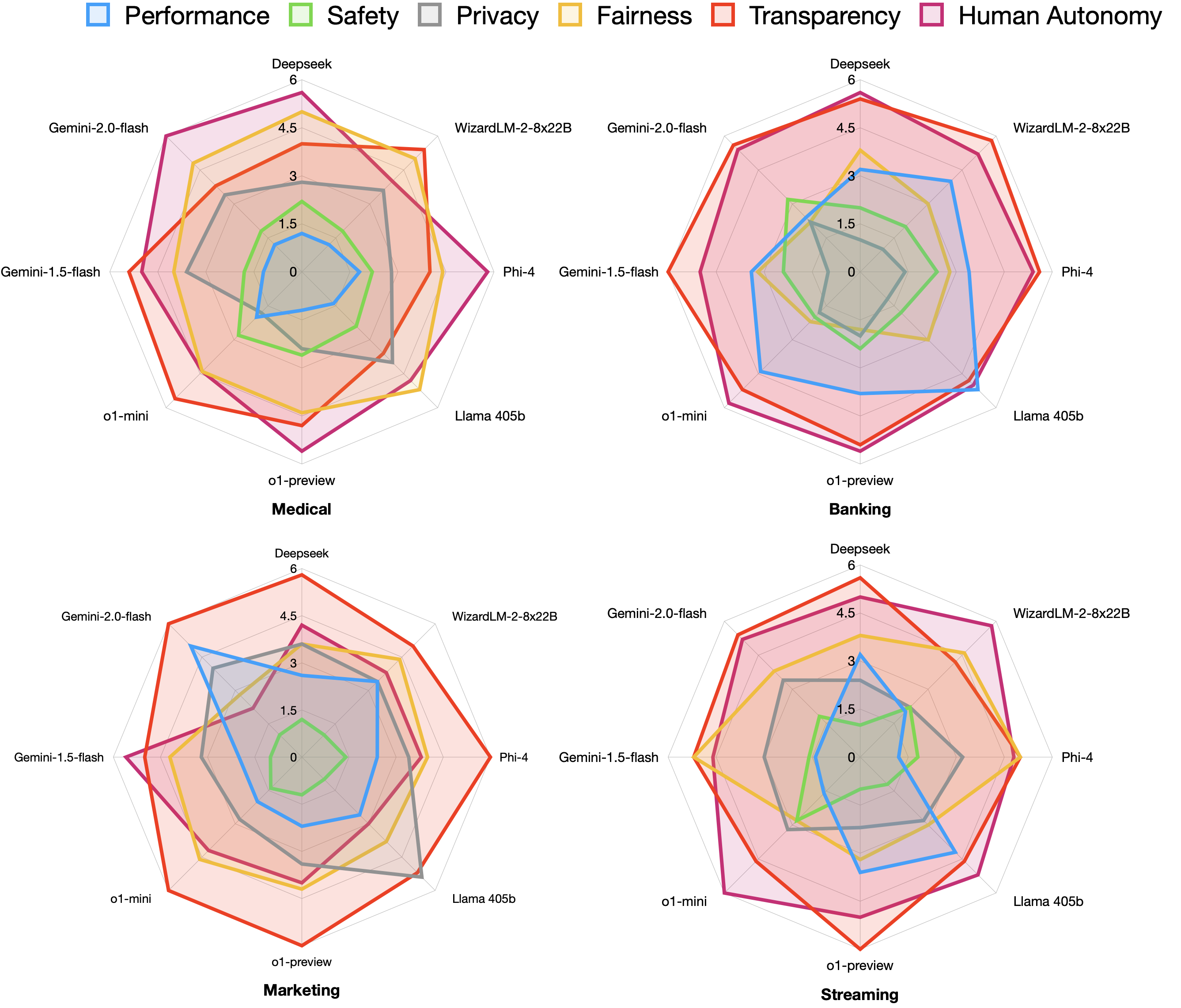} \caption{How LLMs prioritize value-related requirements across different contexts. Lower values indicate higher priority.} \label{fig:Q4} \end{figure*}

We asked LLMs to rank high-level requirements, presented as user stories, to assess how well their priorities align with the reported value profile in tasks T2 and T3. The results are shown in Figure~\ref{fig:Q4}, where lower values indicate higher priority.

In terms of alignment between this task and T2 when prioritizing values in T2 based on how much they were extremely important, we found that 78\% of the ratings (n = 151 out of 192) either aligned (17.7\%) or almost aligned, with a +/- 2 rank difference. The value misalignment occurred the most in safety (n = 14), in which it wasn't deemed extremely important in marketing or streaming by the majority of the models when expressed in abstract terms (T2), then gained high priority when prioritizing high-level requirements (T4). This was followed by transparency (n = 9), in which it was deemed of high priority across models in banking, then had low priority when prioritizing high-level requirements (T4).

Regarding the alignment between this task and T3, we found a strong alignment when it comes to prioritizing safety over human autonomy and safety over transparency in all domains. In addition, both tasks consistently prioritize fairness over performance, especially in the banking context. However, there was less agreement on the trade-offs between privacy, fairness, and performance. In the trade-off task (Figure~\ref{T3}), the responses tended to favor fairness and privacy over performance. In contrast, in this prioritization task, performance was ranked as the highest or second-highest priority in the medical, marketing, and streaming contexts.

\section{Discussion}
\textbf{To what extent are LLMs aligned with human judgment regarding responsible AI values?}

Across the three tasks, LLMs exhibited strong to very strong monotonic positive alignment with AI practitioners, confirmed by statistically significant Spearman correlations (Tables~\ref{T1}-~\ref{T3}). In addition, in task 2, they had a moderate alignment with the US-census representative sample from \cite{Jakesch2022}, and in task 3, they had a very strong alignment with the same US-census sample. LLMs in Group 1 aligned more closely with AI practitioners than with the US-census sample on average. This was especially clear in the first task, where both LLMs and AI practitioners ranked fairness as the most important value, unlike the US-census sample. Additionally, LLMs never ranked performance among their top five values, even though it was the second most important value for the US-census sample. This suggests that LLMs are influenced by the values of AI practitioners, who, on average, did not select performance among the top five values. 

LLMs generally followed human judgment in rating responsible AI values, as shown by the dotted lines in Figure~\ref{fig:Q2byContext}. This indicates that responsible AI values were rated highest in the medical context, followed by banking, marketing, and streaming. LLMs gave a higher average rating for 'extremely important' compared to the combined 'very' and 'extremely important' ratings of the US-census sample in the banking and medical contexts. In the streaming context, although LLMs gave similar or higher importance ratings overall, models like Gemini\_flash 1.5, O1-preview, and Deepseek rated safety, accountability, fairness, and transparency as extremely important in less than 10\% of runs, which is lower than the US-census sample. 

The observed preference for privacy over performance and safety over human autonomy and transparency matches findings from the survey conducted by Jakesch et al.~\cite{Jakesch2022}. Figure~\ref{fig:Q3b} shows a strong correlation between how values are prioritized across contexts by LLMs and the responses from the survey participants, as shown by the dotted bars. For example, although LLMs tend to prioritize privacy and fairness, they prioritize fairness over performance more in the banking context than in others. Similarly, in the performance-privacy trade-off, LLMs prioritize privacy more in marketing than in other contexts.

\emph{These results support the concerns raised by \cite{Jakesch2022} that differences in values between AI practitioners and end users may influence AI models, such as LLMs, to reflect the values of the developers rather than the users.} While no LLM perfectly aligns with the US-census sample nor AI practitioners, based on Tables \ref{T2} and \ref{T3}, the O1-preview model is more aligned with AI practitioners, followed by Llama 3.1 405b. On the other hand, Gemini-2.0-flash is more aligned with the US-census sample, followed by O1-mini.
\par

\textbf{What Characteristics of LLM Models Influence Their Stated Value Preference?}
While many factors can influence the stated value preference of LLMs, we focused on three key characteristics: size, release order, and family, as these can be directly observed from the tasks outlined.

\textbf{Size.} From our preliminary experiments on model robustness, we found that smaller models fall into Group 3, observed in Llama and Gemma LLMs, showing higher variability in the value selection task, indicating that their choices were almost random. Regarding alignment with the US-representative sample and AI practitioners. Observed in T2,  larger LLMs tend to have higher correlations with AI practitioners than smaller counterparts. However, this trend is not statistically significant due to the small sample size. Additionally, this pattern is not seen in Task 3, where O1-preview was more aligned with the US-representative sample than O1-mini.

\textbf{Release Order.} We observed that different releases within the same LLM family behaved differently in terms of alignment with the US-representative sample and AI practitioners. OpenAI models in the O1 release tended to align more with AI practitioners and less with the US-representative sample. In contrast, Gemini and Llama models showed the opposite trend. 

\begin{figure} \centering \includegraphics[width=0.9\linewidth]{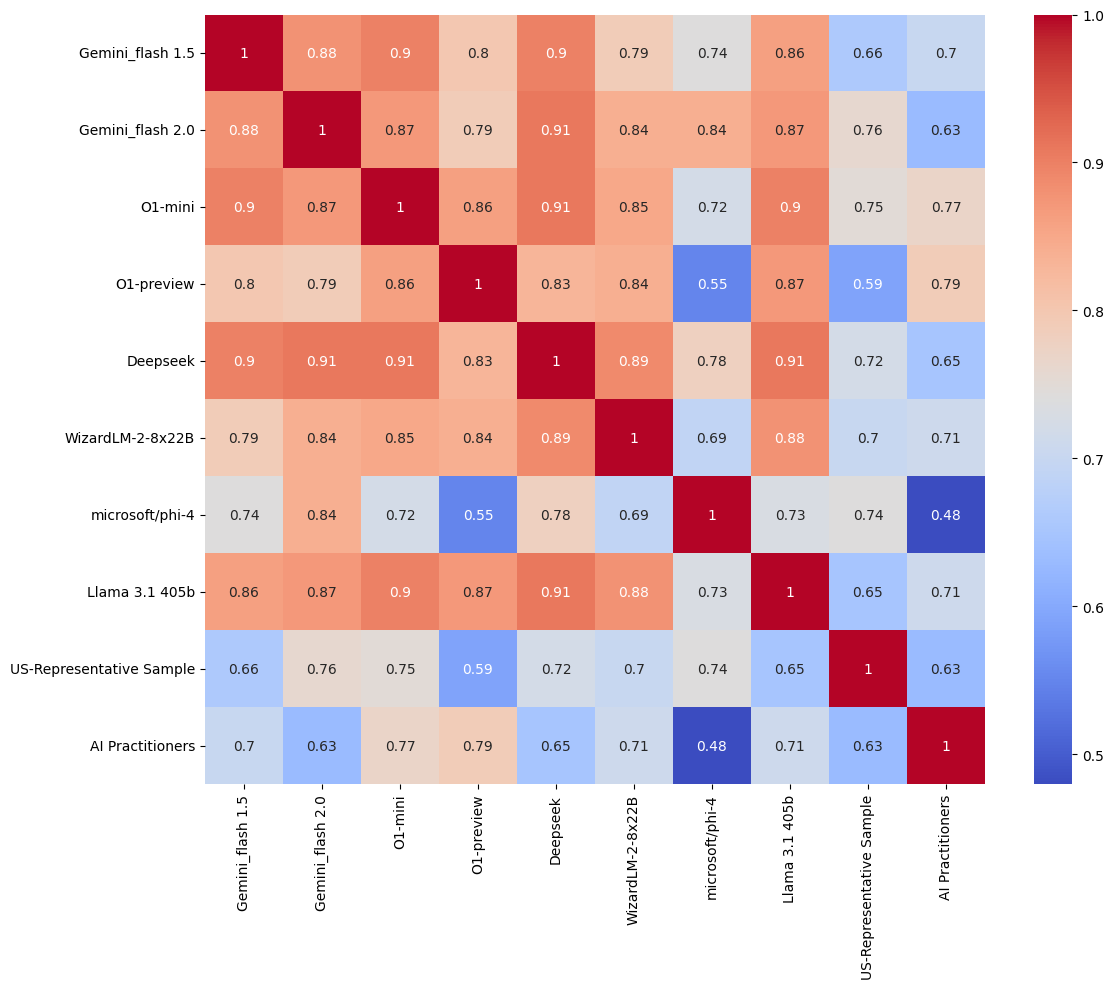} \caption{Spearman correlations between LLMs in Group 1.} \label{fig:heat} \end{figure}

\textbf{LLM Family.} Figure~\ref{fig:heat} shows a heatmap of Spearman correlations between different LLMs based on T2. Although models from the same family show strong correlations with each other, they can also have high correlations with models from other families.

\textit{Based on the 23 LLMs from various families, investigated models with fewer than 70 billion parameters show signs of lacking the reasoning capability needed for ethical decision-making. We can conclude that an LLM's value system cannot be predicted based solely on its size, family, or release order.
}
\textbf{Do LLMs' requirements prioritization behaviors align with their stated preferences?}
We explored whether LLMs can be effectively used in requirements engineering tasks while ensuring that their outcomes align with the value profiles identified in T2 and T3. The added requirements prioritization task highlighted practical discrepancies between LLMs' stated abstract value preferences and their actual decision-making behavior, offering deeper insight into their alignment robustness. The results showed that although simple preference questions provide a general understanding of an LLM's value inclinations, they cannot strictly translate to $value A \le value B$. Values such as privacy, fairness and performance are generally considered highly important as abstract concepts (T2) and in requirements prioritization over human autonomy (T4). However, in a consistent manner, LLMs are indecisive or highly uncertain when choosing between privacy, fairness, and performance for T3-T4, even sometimes within the same LLM in different runs. Hence, equal importance in T2 or high uncertainty in T3 leads directly to misalignment or an inability to prioritize in T4. This was observed more with the value of fairness. Although fairness was identified as the most important value in the abstract value rating (T1), its priority is reduced during concrete decision-making tasks in T4.

\textbf{What is the source of misalignment between prioritization behaviors align with their stated preferences?}
In this preliminary investigation, we identified two types of misalignment: (1) between LLMs and the general public, and (2) between LLMs’ stated preferences (Tasks 1–3) and their behavior in requirements prioritization (Task 4).

For the first type, our results show that LLMs align more closely with AI practitioners than with a US-representative sample. Although it is hard to pinpoint the exact reason, several well-documented phenomena might be the source of misalignment. 1- Training data and system design with cultural values of the US-based development team getting embedded into the models~\cite{tao2024cultural}. Similarly, ~\cite{Jakesch2022} demonstrates that practitioners and end users prioritize responsible AI values differently, and our findings suggest that such practitioner influence transfers into model outputs. 2- Annotation and feedback pipelines used in Reinforcement Learning from Human Feedback (RLHF) could be prone to selection bias by the researchers and often drawn from skilled annotators or domain experts, reinforcing practitioner perspectives over general-population views~\cite{casper2023openproblems}.

As for the second type, it echoes prior findings that LLMs often struggle with faithfulness—maintaining consistency between abstractly stated positions and applied behavior~\cite{turpin2023languagemodelsdontsay}. For instance, ~\cite{agarwal2024faithfulnessvsplausibilityunreliability,turpin2023languagemodelsdontsay} show that models can produce unfaithful explanations or inconsistent reasoning chains, revealing a gap between what models say and what they do. Although we cannot pinpoint for sure what causes such inconstancy, ~\cite{santurkar2023opinionslanguagemodelsreflect} emphasized that steering the model to certain values could cause inconsistency. Aside from lack of faithfulness, it could be due to the LLMs following other than dominant aspect prioritization method such as risk reduction via infringement \& amelioration, and weighted scores~\cite{p44}.

\textbf{Implications.} The findings highlight critical considerations for both requirements engineers and AI developers when deploying LLMs in software engineering workflows. 
Although recent studies advocate using LLMs for requirements elicitation, analysis, and prioritization~\cite{sami2024experimentingmultiagentsoftwaredevelopment,sami2024prioritizing,10.1145/3695988}, our results demonstrate that such use demands caution. 
Even advanced models exhibit inconsistent alignment with responsible AI values, as shown in Task 4, often producing unstable prioritization or conflicting trade-off judgments—echoing earlier observations with GPT-3.5~\cite{Norhashim_Hahn_2024}. 
These inconsistencies underscore the need for more rigorous evaluation and probing techniques than those currently applied in value-assessment studies such as~\cite{Jakesch2022}.

In practice, currently, LLMs should serve as \textit{assistive tools} only and
requirements engineers must remain actively involved in prompting, tuning for value alignment, and reviewing model outputs to ensure ethically grounded outcomes. 
Interestingly, the observed alignment between LLMs and AI practitioners—particularly around fairness, transparency, and accountability—suggests that engineering choices influence ethical orientation. 
Expanding feedback sources and incorporating perspectives from broader demographic and cultural groups could therefore foster more inclusive value alignment. 
Achieving such alignment will require multidisciplinary collaboration between AI developers, requirements engineers, and end users to systematically embed diverse ethical priorities into LLM-assisted development.


\textbf{Limitations}
Limitations of this study include: (1) Reliance on~\cite{Jakesch2022} for the US census-representative sample's position on responsible AI value which represents the US population only, hence it does not necessarily generalize to the population of other countries with different value systems. (2) The~\cite{Jakesch2022} study was conducted in 2022 and the public's perception of what values are relevant could change. (3) Limited ablation study regarding the different hyper-parameters and variations related to the LLMs, as we only model size and release order within an LLM family. (4) Our study only evaluates the LLMs when prompted in English only. (5) Inherent stochasticity in LLM outputs may still introduce variability even if a certain LLM showed high robustness in the experiments. (6) Although high alignments was observed in LLMs with AI practitioners, this does not necessarily reflect the output when confronted with a real-world more complex scenario. (7) Evaluations based on requirements prioritization tasks in four contexts might not fully capture the range of values influencing LLM behavior, particularly regarding faithfulness concerns. Results may not generalize to other requirements engineering tasks or even to different systems within the same contexts.

\section{Conclusion}
This study examined the alignment of LLMs with human values related to responsible AI, comparing their judgments with those of a US-representative sample and AI practitioners. Our analysis across three tasks showed that LLMs generally aligned more closely with AI practitioners, particularly emphasizing fairness, privacy, transparency, safety, and accountability. Although no LLM perfectly aligns with either the US-representative sample or AI practitioners, the O1-preview model appeared most aligned with AI practitioners, followed by Llama 3.1 405b, whereas Gemini-2.0-flash was more aligned with the US-representative sample, followed by O1-mini. However, alignment was inconsistent across different values and varied significantly depending on context and specific value conflicts.

These findings suggest that while LLMs do reflect the values of their developers and the broader AI community, they may not fully represent the diverse range of values held by the general public. This discrepancy underscores the importance of carefully evaluating and guiding the integration of LLMs into applications that impact varied user groups. Furthermore, inconsistencies observed in value prioritization tasks indicate that relying solely on LLMs for requirements engineering and value prioritization is currently insufficient.

Future research should investigate methods to explicitly align LLMs with prioritized human values specific to targeted demographic groups. Additionally, in-depth cross-model comparisons regarding different responsible AI value profiles should be conducted. Studies should also expand beyond US-centric samples to include diverse populations globally. While LLMs hold significant potential to advance responsible AI development, their integration into human-centric processes requires ongoing attention and refinement. Establishing comprehensive benchmarks across various requirements engineering tasks is essential to enable more robust validation of LLM alignment.

\section*{Acknowledgment}
The authors would like to acknowledge the support received from Saudi Data and AI Authority (SDAIA) and King Fahd University of Petroleum and Minerals (KFUPM) under SDAIA-KFUPM Joint Research Center for Artificial Intelligence Grant no. JRC-AI-RG-12.

\bibliographystyle{ieeetr}
\bibliography{sample-base}

\end{document}